\documentclass[12pt]{iopart}
\usepackage{iopams}
\usepackage{bbold}

\DeclareFontFamily{U}{bbold}{}
\DeclareFontShape{U}{bbold}{m}{n}
   {  <5> <6> <7> <8> <9> gen * bbold
      <10> <10.95> bbold10
      <12> <14.4> bbold12
      <17.28> <20.74> <24.88> bbold17
   }{}

\newcommand{\eqref}[1]{(\ref{#1})}
\newcommand{\labl}[1]{\label{#1}}
\newcommand{\equ}[1]{\begin{eqnarray} #1 \end{eqnarray}}
\newcommand{\equa}[1]{\begin{eqnarray} #1 \end{eqnarray}}
\renewenvironment{matrix}{%
  \hskip -\arraycolsep\array{cccccccccc}%
}{%
  \endarray \hskip -\arraycolsep
}
\renewenvironment{pmatrix}{\left(\matrix}{\endmatrix\right)}
\newcommand{\pmtrx}[1]{\begin{pmatrix} #1 \end{pmatrix}}
\newcommand{\non}{\nonumber}

\newcommand{\vc}[1]{\bi{#1}}
\newcommand{\mx}[1]{\mathbf{#1}}
\renewcommand{\d}{\rmd}
\newcommand{\ml}{\ll}
\DeclareMathSymbol{\mg}{\mathrel}{symbols}{"1D}     
\newcommand{\dd}[2][]{\ensuremath{\frac{\d #1}{\d #2}}}

\newcommand{\Id}{\mathbb{1}}
\newcommand{\Bnabla}{\bnabla}

\newcommand{\der}{\partial}
\newcommand{\inv}{^{-1}}
\newcommand{\lh}{\left(}
\newcommand{\rh}{\right)}
\newcommand{\half}{\frac{1}{2}}

\newcommand{\ga}{\alpha}

\renewcommand{\gg}{\gamma}
\newcommand{\gd}{\delta}

\newcommand{\gf}{\phi}

\newcommand{\gm}{\mu}

\newcommand{\gk}{\kappa}

\newcommand{\gr}{\rho}

\newcommand{\gG}{\Gamma}
\newcommand{\gD}{\Delta}

\newcommand{\gL}{\Lambda}

\newcommand{\gO}{\Omega}

\newcommand{\cD}{{\cal D}}

\newcommand{\cH}{{\cal H}}

\newcommand{\cL}{{\cal L}}

\newcommand{\tn}{{\tilde n}}

\newcommand{\tM}{{\tilde M}}

\newcommand{\tS}{{\tilde S}}

\newcommand{\tge}{{\tilde\epsilon}}

\newcommand{\tgx}{{\tilde\xi}}

\newcommand{\tget}{{\tilde\eta}}

\newcommand{\Bgd}{\bdelta}

\newcommand{\Bgf}{\bphi}

\newcommand{\Bgx}{\bxi}

\newcommand{\Bget}{\bfeta}

\begin{document}


\title{Multiple-field inflation and the CMB}

\author{Bartjan van Tent}

\address{Institute for Theoretical Physics, University of Amsterdam,
Valckenierstraat~65, 1018~XE~Amsterdam, The~Netherlands}

\ead{bvtent@damtp.cam.ac.uk}

\begin{abstract}
In this paper, we investigate some consequences of multiple-field inflation for
the cosmic microwave background radiation (CMB).
We derive expressions for the amplitudes, the spectral indices and the
derivatives of the indices of the CMB power spectrum in the context of a very
general multiple-field theory of slow-roll inflation, where the field metric can
be non-trivial. Both scalar (adiabatic, isocurvature and mixing) and tensor
perturbations are treated and the differences with single-field inflation are
discussed. From these expressions, several relations are derived that can be 
used to determine the importance of multiple-field effects observationally from 
the CMB. We also study the evolution of the total entropy perturbation during
radiation and matter domination and the influence of this on the isocurvature 
spectral quantities.
\end{abstract}


\section{Introduction}

One of the greatest successes of the concept of inflation \cite{Guth,
Starobinskymodel,boekLinde,LythRiotto,LiddleLythboek} is that it can give an
explanation for the existence  of small density fluctuations in an otherwise
homogeneous universe. These small fluctuations, which are the gravitational
seeds for the formation of large-scale structures, are observed in the cosmic
microwave background radiation (CMB). Of course this works two ways:
the observed amplitudes and slopes of the fluctuation spectra give us some
observational constraints on the otherwise rather elusive parameters in
inflation models, and thus on the parameters in the underlying high-energy
theories. Hence it is important to have expressions for these spectral
quantities in terms of the parameters of the inflation models. The
most accurate observations of the CMB to date are those from the Wilkinson
Microwave Anisotropy Probe (WMAP) mission \cite{MAP} (see 
\cite{WMAP,WMAPparam,WMAPinfl}).

This paper is a sequel to our paper \cite{GNvT2}. In that paper a general 
theory of inflation, with an arbitrary number of scalar fields that take values
on a possibly non-trivial field manifold, was considered (motivated by
(string-inspired) supergravity theories where the K\"ahler potential leads to
non-minimal kinetic terms). We treated both the background, paying special 
attention to the concept of slow roll, and the scalar perturbations. 
After deriving expressions valid to first order in slow roll for
the perturbation quantities at the end of inflation, we also considered
what happens after inflation. This finally led to expressions for the
correlators of the adiabatic and isocurvature scalar perturbations, as well as
of the mixing between them, at the time of recombination (basically these are 
the scalar amplitudes of the CMB power spectrum). An example of a quadratic 
potential was included to illustrate the theory.

In this paper, these results are extended in basically two ways. In the first
place, expressions for the spectral indices $\tn$ and their derivatives  
$\d\tn / \d \, \ln k$ are given next to those for the amplitudes, and results 
for the tensor perturbations are added as well. Using these results we derive 
an expression that can be used to observationally test the importance of 
multiple-field effects in general, as well as various consistency relations for 
the one- and two-field cases. Secondly, the behaviour of the total entropy 
perturbation after inflation is studied to investigate under which conditions 
the usual assumption of a constant isocurvature perturbation is justified and 
what happens to the isocurvature spectral quantities if this is not the case.

A large number of references to other papers on inflationary perturbation 
theory was given in the introduction of \cite{GNvT2}. More recently, a 
second-order treatment of the single-field case was given in
\cite{Acquavivaetal}, while the authors of \cite{Habibetal,MartinSchwarzWKB}
used non-slow-roll techniques in the single-field case. Many aspects of the
two-field case were recently discussed in \cite{LesgourguesPolarski,Bartoloetal,
Wandsetal2, Tsujikawaetal} (especially the last of these contains a very large 
number of references to earlier papers). An extensive quantum treatment, 
including backreaction, of O(N)-symmetric multiple-field inflation models was 
given in \cite{BoyanovskyDeVega}.
Previous work on consistency relations for two-field inflation can be found in
\cite{PolStargrav,SasakiStewart,GarciaBWands,Bartoloetal,Wandsetal2}.

The outline of the paper is as follows. Section~\ref{multisec} deals with the
general theory of multiple-field inflation. In three subsections the background,
scalar perturbations and tensor perturbations are discussed. Most of this
(except for the tensor perturbations) was derived in \cite{GNvT2}, so that this 
section is kept brief, only setting up the notation and giving the results that 
are used in the rest of the paper. Derivations are only given for those results 
that were not present in \cite{GNvT2}. 
In section~\ref{corrinfl} (plus the appendix), expressions for the scalar 
(adiabatic, isocurvature and mixing) and tensor amplitudes, spectral indices, 
and the derivatives of the spectral indices are derived. The results are 
discussed with special attention to the differences with single-field
inflation. In subsection~\ref{obstestsec}, a number of important relations
between these quantities  are derived that are relevant for observationally
testing the significance of multiple-field effects.
Section~\ref{entropypertsec} is devoted to the total entropy perturbation,
which enters as a source term in the equation of motion for the gravitational
potential after inflation in a system with multiple components originating from
multiple fields during inflation, and thus determines the isocurvature 
perturbation. An equation for its time derivative is derived  and the resulting
time dependence is discussed. Finally, section~\ref{obsconcl} contains the
conclusions. Some of the results in this paper have appeared in \cite{thesis}.

\section{Multiple-field inflation}
\labl{multisec}

After defining the general setup of the theory, the results for all the 
relevant perturbation quantities at the time of recombination are given in 
this section. For the derivation of the scalar results as well as for more
details about the background quantities the reader is referred to
our previous paper \cite{GNvT2}; the derivation of the tensor results is
given here. Readers interested in seeing more intermediate steps in the
calculations might also want to take a look at \cite{thesis}. 

\subsection{Background}

We consider the situation where the matter content of the universe consists of
an arbitrary number of real scalar fields $\gf^a$, which are represented as a 
vector $\Bgf$ of scalar field components. These fields are the coordinates on a
possibly non-trivial field manifold with field metric $\mx{G}$. The field is
separated into a classical homogeneous background part and a small quantum
perturbation part: 
$
\Bgf^{\mathrm{full}}(\eta,\vc{x}) = \Bgf(\eta) + \Bgd\Bgf(\eta,\vc{x}).
$
All equations are linearized with respect to the perturbations.
The associated Lagrangean with potential $V$ is
\equ{
\fl
\cL = 
\sqrt{-g} \lh - \half \der^\gm \Bgf \cdot \der_\gm\Bgf - V(\Bgf) \rh
\equiv
\sqrt{-g} \lh - \half g^{\mu\nu} \der_\mu \Bgf^T \mx G \der_\nu \Bgf
- V(\Bgf) \rh,
\labl{Lagrangean}
}
where $T$ denotes the transpose and $g_{\mu\nu}$ is the spacetime metric.
Because of the non-trivial field metric we have to define covariant derivatives 
instead of normal ones:
$
\nabla_b A^a \equiv A^a_{\;\: ,b} + \gG^a_{bc} A^c
$
and
$
\cD_\mu A^a \equiv \der_\gm A^a + \gG^a_{bc} \der_\gm \gf^b A^c
$
for the field derivative and the spacetime derivative of a vector $A^a$ in field
space.

For the metric part of the universe we make the following definition:
\equ{
\fl
g_{\mu\nu} 
= a^2 \pmtrx{-1 & 0\\ 0 & \gd_{ij}}
- a^2 \pmtrx{2\Phi & 0\\ 0 & 2 \Phi\gd_{ij}}
+ a^2 \pmtrx{0 & S_j\\ S_i & 0}
+ a^2 \pmtrx{0 & 0\\ 0 & h_{ij}},
\labl{fullmetric}
}
where we have applied the longitudinal and vector gauges and used the
($ij$)-component ($i \neq j$) of the Einstein equation to set $\Psi=\Phi$ in 
the scalar part (see e.g.\ \cite{Mukhanovetal,Stewart}). 
The first term represents the homogeneous
background (a flat Robertson-Walker metric with scale factor $a(\eta)$), the
second term the scalar perturbation (with $\Phi(\eta,\vc{x})$ the gravitational
potential), the third term the vector perturbation (with $\vc{S}(\eta,\vc{x})$ a
divergenceless vector), and the fourth term the tensor perturbation
(with $h_{ij}(\eta,\vc{x})$ a symmetric transverse traceless tensor, i.e.\
$h_{ij}=h_{ji}$, $h^{ij}_{\;\: ,i}=0$ and $h^i_{\; i}=0$).
As proved in \cite{Bardeen,Stewart} scalar, vector and tensor perturbations
decouple to linear order and can be considered separately.
We make use of three different time coordinates: comoving time $t$, conformal
time $\eta$ and the number of e-folds $N$, defined as
\equ{
\fl
\d t = a \, \d\eta,
\qquad\quad
\d N = H \d t = \cH \d\eta,
\qquad\quad
H \equiv \frac{\dot{a}}{a},
\qquad\quad
\cH \equiv \frac{a'}{a} = a H,
}
where $H$ and $\cH$ are the Hubble parameters associated with $t$ and $\eta$,
respectively, and a dot (prime) denotes a derivative with respect to comoving
(conformal) time.

For giving a physical interpretation of the various scalar field components, it
is very useful to define a basis on the field manifold. (It is also an important
ingredient of the calculations, especially quantization, as explained in
\cite{GNvT2}, where we introduced this basis.) We define the first basis vector
$\vc{e}_1$ as the direction of the field velocity: 
$\vc{e}_1 \equiv \dot{\Bgf}/|\dot{\Bgf}|$. Next, $\vc{e}_2$ is the direction of 
that part of the field acceleration $\cD_t \dot{\Bgf}$ that is perpendicular to 
$\vc{e}_1$, and this is extended to higher-order derivatives to define the 
other basis vectors. Once this basis is defined we can use projection operators
to split vectors into their physical components. The most important ones are 
$\mx{P}^\parallel \equiv \vc{e}_1^{\,} \, \vc{e}_1^T \mx{G}$ and 
$\mx{P}^\perp \equiv \Id - \mx{P}^\parallel$ that make it possible to
distinguish between effectively single-field ($\vc{e}_1$) and truly
multiple-field effects. The basis vectors are not constant in time; their time
derivatives are given by
\equ{
\cD_t \vc{e}_n
= H \lh \frac{\tget^{(n+1)}_{~n+1}}{\tget^{(n)}_{~n}} \, \vc{e}_{n+1} 
- \frac{\tget^{(n)}_{~n}}{\tget^{(n-1)}_{~n-1}} \, \vc{e}_{n-1} \rh,
\labl{basisder}
}
where $\tget^{(n)}_{~n} \equiv \vc{e}_n \cdot \tilde{\Bget}^{(n)}$, with
$\tilde{\Bget}^{(n)}$ defined below. 

The multiple-field generalization of the single-field slow-roll approximation
was treated in \cite{GNvT2} as well. We defined the following slow-roll 
functions:
\equ{
\tge \equiv - \frac{\dot{H}}{H^2} = \frac{\half \gk^2 |\dot{\Bgf}|^2}{H^2},
\qquad\qquad
\tilde{\Bget}^{(n)} \equiv 
\frac {\cD_t^{n-1} \dot{\Bgf}}{H^{n-1} |\dot{\Bgf}|}, 
\labl{slowrollfun}
}
and the short-hand notation $\tilde{\Bget} \equiv \tilde{\Bget}^{(2)}$ and 
$\tilde{\Bgx} \equiv \tilde{\Bget}^{(3)}$. Here $\gk$ is the inverse reduced 
Planck mass: $\gk^2 \equiv 8\pi G = 8\pi/M_P^2$.
Using the basis discussed above we can
take the components $\tget^\parallel \equiv \vc{e}_1 \cdot \tilde{\Bget}$ and 
$\tget^\perp \equiv \vc{e}_2 \cdot \tilde{\Bget}$ (by construction there are no 
other components), and similarly for $\tilde{\Bgx}$, although there one has not 
just a single perpendicular direction, but two ($\tgx_2$ and $\tgx_3$).
Using these definitions the equations of motion for the background quantities
are:
\equ{
\fl
H = \frac{\gk}{\sqrt{3}} \, \sqrt{V} \lh 1 - \frac{1}{3} \tge \rh^{-1/2},
\qquad
\Bgf_{,N} + \frac{1}{\gk^2} \frac{\mx{G}\inv \Bnabla^T V}{V}
= - \frac{\sqrt{2}}{3\gk} \frac{\sqrt{\tge} (\tilde{\Bget} + \tge\, \vc{e}_1)}
{1-\frac{1}{3}\tge},
\labl{eqmotbackN}
}
where we have switched to using the number of e-folds $N$ as time coordinate
because it simplifies the equation.
These equations are still exact; the slow-roll functions have only been used as
short-hand notation. However, the next step is the slow-roll approximation: 
we assume that the slow-roll
functions are small so that we can set up a series expansion in terms of them.
($\tge$ and $\tilde{\Bget}$ are assumed to be first-order quantities, while
$\tilde{\Bget}^{(n)}$ is of order $n-1$.)
Note that the time derivatives of the slow-roll functions are of one order
higher in slow roll:
\equa{
\fl
\tge_{,N} = 2 \tge ( \tge + \tget^\parallel ), 
&
\cD_N \tilde{\Bget}^{(n)} = \tilde{\Bget}^{(n+1)} + 
( (n-1) \tge - \tget^\parallel ) \tilde{\Bget}^{(n)},
\non\\
\fl
(\tget^\parallel)_{,N} = \tgx^\parallel 
+ (\tget^\perp)^2 + (\tge - \tget^\parallel) \tget^\parallel,
\qquad &
(\tget^\perp)_{,N} = \tgx_2 + (\tge - 2\tget^\parallel) \tget^\perp. 
\labl{dereps}
}

\subsection{Scalar perturbations}
\labl{scalarpertsec}

The derivation of expressions for the scalar perturbations was the main
topic of \cite{GNvT2}. Because the equations of motion for the gravitational
potential and the field perturbations do not decouple in the multiple-field
case, we cannot ignore the latter as one can in the single-field case.
The scalar-field perturbations are redefined and quantized as
\equ{
\vc{q} \equiv a \lh \Bgd\Bgf + \frac{\Phi}{\cH} \, \Bgf' \rh,
\qquad\qquad
\hat q = Q \hat a^\dag + Q^* \hat a,
\labl{defqwithR}
}
with $\hat{a}^\dag$ and $\hat{a}$ constant creation and annihilation operator
vectors and $Q(\eta)$ a matrix function that satisfies a classical equation of
motion. Here we switched to working in the explicit basis: non-bold versions of 
a vector or matrix represent those quantities in this basis, for example,
$
q^T = (q_1,q_2,\ldots) = 
(\vc{e}_1 \cdot \vc{q}, \vc{e}_2 \cdot \vc{q}, \ldots).
$
This means that the non-bold basis vectors are simple constant vectors: 
$e_1^T = (1,0,0,0,\ldots)$, $e_2^T = (0,1,0,0,\ldots)$, etc.
In the rest of this section we will denote expressions like the one for
$\hat{q}$ in \eqref{defqwithR} as $\hat{q} = Q \hat{a}^\dag + \mbox{c.c.}$, with
the c.c.\ meaning complex conjugate.

The results for the adiabatic and isocurvature parts (see
below for a definition as well as for some remarks about assumptions made in the
derivation) of the gravitational potential in terms of Fourier modes
at the time of recombination, valid to first order in slow roll for modes with
$k\ml\cH_{\mathrm{rec}}$ (super-horizon modes), are
\equa{
\hat{\Phi}_{\vc{k}\, \mathrm{ad}}(t_{\mathrm{rec}}) 
= & \frac{3}{5} \frac{\gk}{2 k^{3/2}} \frac{H_\cH}{\sqrt{\tge_\cH}}
\lh e_1^T + U_{P\, e}^T \rh E_\cH
\hat{a}_{\vc{k}}^\dag + \mbox{c.c.},
\labl{Phiadafter}\\
\hat{\Phi}_{\vc{k}\, \mathrm{iso}}(t_{\mathrm{rec}}) 
= \frac{1}{6} \, & 
\frac{3}{5} \frac{\gk}{2 k^{3/2}} \frac{H_\cH}{\sqrt{\tge_\cH}}
V_{e}^T E_\cH \hat{a}_{\vc{k}}^\dag + \mbox{c.c.}
\labl{Phiisoafter}
}
Here we have defined
$E_\cH \equiv (1- \tge_\cH) \Id  + (2 - \gg - \ln 2) \gd_\cH$, with $\gg$ the 
Euler constant, and\footnote{For normalization reasons, we have removed a factor 
$\half$ from the definition of $V_e$ as compared with our original definition 
in \cite{GNvT2}.} 
\equa{
\fl
\gd \equiv \tge \, \Id - \frac {\tM^2}{3H^2} + 2 \tge \, e_1^{\;} e_1^T,
& 
\tilde{\mx M}^2 \equiv \mx G\inv \Bnabla^T \Bnabla V 
- \mx R (\dot{\Bgf},\dot{\Bgf}),
\labl{defdelta}\\
\fl
U_{P\, e}^T \equiv 2 \sqrt{\tge_\cH} \int^{t_e}_{t_\cH} \d t' \, H
\frac{\tget^\perp}{\sqrt{\tge}} \, \frac{a_\cH}{a} \, e_2^T Q Q_\cH\inv,
\qquad & 
V_e^T \equiv \sqrt{\tge_\cH} \, \frac{\sqrt{\tge_e^{\;}} \, \tget^\perp_e}
{\tge_e + \tget^\parallel_e} \frac{a_\cH}{a_e} \, e_2^T Q_e^{\;} Q_\cH\inv.
\labl{defUPe}
}
The subscript $\cH$ means that the quantity has to be evaluated at the time
$t_\cH$ during inflation when $\cH = k$ (often called the time of horizon 
crossing). The quantity $Q_\cH$ is given by $Q_\cH = E_\cH/\sqrt{2k}$.
Note that the slow-roll approximation was only used during a small
transition region around the time $t_\cH$. The time $t_e$ is the end of 
inflation. The vectors $U_{P\, e}$ and $V_e$ have no components in the $e_1$ 
direction. The matrix $\mx{R}$ is the curvature tensor on the field manifold and 
$[\mx{R}(\dot{\Bgf},\dot{\Bgf})]^a_{\; d} = R^a_{\; bcd}\dot{\gf}^b\dot{\gf}^c$. 

The terms adiabatic and isocurvature relate to the evolution after inflation.
During radiation and matter domination the equation of motion for super-horizon
modes of $\Phi$,
\equ{
\ddot{\Phi} + 4 H(1+{\textstyle\frac{3}{4}}c_s^2) \dot{\Phi} 
+ \gk^2 (\gr c_s^2 - p) \Phi = - 2 \gk^2 (\gr c_s^2 - p) \tS,
\labl{Phiafter}
}
has an inhomogeneous source term proportional to the total entropy 
perturbation $\tS$:
\equ{
\tS \equiv \frac{1}{4} \frac{\gd p - c_s^2 \gd\gr}{p - c_s^2 \gr},
\labl{defspgr}
}
with $p$ and $\gr$ the total pressure and energy density and the sound velocity
$c_s^2 \equiv \dot{p}/\dot{\gr}$. 
The adiabatic perturbation is the homogeneous solution for~$\Phi$ of this
equation of motion, while the isocurvature perturbation is the particular
solution generated by the source term, with the initial conditions that it is 
zero and has zero derivative at the beginning of the radiation-dominated era. 
If $\tS$ is constant on super-horizon scales, we have the simple solution 
$\Phi_{\mathrm{iso}}=-\frac{1}{5}\tS$ during the matter era, 
which was used in the derivation of equation \eqref{Phiisoafter}.
To perform the matching of these solutions with the ones at the end of inflation
(to determine the constants of integration) we assumed an immediate transition
to a radiation-dominated universe at the end of inflation, ignoring 
(p)reheating. Especially for the isocurvature perturbations this is expected to
be a crude approximation (since $V_e$ depends crucially on what happens at the
end of inflation, see \eqref{defUPe}), which needs further improvement, 
but the treatment of the perturbations during a more realistic transition at 
the end of inflation as well as during an epoch of (p)reheating is still under 
investigation.\footnote{Some study on 
the effects of preheating on the scalar perturbations has been done for 
specific models, see \cite{Bassettetal,Wandsetal,FinelliBrandenberger,
thesisMalik,HenriquesMoorhouse,TsujikawaBassett,TanakaBassett} and references 
therein, but different authors do not yet agree regarding the conclusions.}

The behaviour of $\tS$ during radiation and matter domination is investigated 
in section~\ref{entropypertsec}, to see if the assumption of constancy on 
super-horizon scales is justified. However, let us discuss here what happens 
with \eqref{Phiisoafter} in the case that $\tS$ is not constant, but given by
$\tS(t) = f(t) \tS_e$. Here $f(t)$ encodes both the possible time dependence of 
$\tS$ during radiation and matter domination and a possible correction factor 
to the simple matching condition described above, and $\tS_e$ is the total 
entropy perturbation at the end of inflation, given by \cite{GNvT2,thesis}
\equ{
\tS_e = \frac{\gk}{2\sqrt{2}} \left [ \frac{\sqrt{\tge}}{\tge+\tget^\parallel} 
\, \tget^\perp \, \frac{q_2}{a} \right ]_e. 
\labl{tSinfl}
}
A particular solution for the inhomogeneous equation \eqref{Phiafter} is now 
given by $\Phi_{\mathrm{part}}(t) = - 2 F(t) \tS_e$, where ($w \equiv p/\gr$)
\equ{
F(t) \equiv \frac{H}{a} \int^t \d t' 
\frac{1+w}{\frac{5}{6}+\half w} \, \dd{t'} \lh \frac{a}{H} \rh
\int^{t'} \d t'' \, f(t'') \, \dd{t''} \lh \frac{\frac{5}{6}+\half w}{1+w} \rh.
\labl{defFt}
}
Adding the homogeneous solution 
$\Phi_{\mathrm{hom}} = C (H/a) + D (H/a) \int^t \d t' (1+w) a$,
with appropriate constants $C$ and $D$ to satisfy the initial conditions,
we finally obtain the new result for $\Phi_{\mathrm{iso}}(t_{\mathrm{rec}})$.
It turns out that we can represent it by exactly the same expression as in 
equation \eqref{Phiisoafter}, if we change the definition of $V_e$ to
\equ{
V_e^T \equiv \left [ 10 F(t_{\mathrm{rec}}) - 3 \lh 3 F(t_*)
+ \frac{\dot{F}(t_*)}{H_*} \rh \right ]
\sqrt{\tge_\cH} \, \frac{\sqrt{\tge_e^{\;}} \, \tget^\perp_e}
{\tge_e + \tget^\parallel_e} \frac{a_\cH}{a_e} \, e_2^T Q_e^{\;} Q_\cH\inv,
\labl{defVe2}
}
with $t_*$ the beginning of the radiation-dominated era.
Note that $V_e$ has now lost the physical interpretation of a quantity
completely determined by inflation, but mathematically it is very convenient
to work with an expression for $\Phi_{\mathrm{iso}}(t_{\mathrm{rec}})$ in which
only $V_e$ changes, as opposed to changing the form of \eqref{Phiisoafter} while
keeping the definition of $V_e$ fixed.
In the case that $f(t)=1$ for all $t$, one finds $F(t)=1$ for all $t$,
and \eqref{defVe2} reduces to \eqref{defUPe}, as it should.

\subsection{Tensor perturbations}
\labl{tensorpert}

Having reviewed the scalar perturbation results from \cite{GNvT2} in the 
previous subsection, we now turn to tensor perturbations. Scalar-field
perturbations cannot, by definition, generate tensor perturbations in the
metric. However, the two tensor degrees of freedom of the metric are the only
physical ones (representing the two polarizations of the graviton) and they do
not need to be generated by a matter source.\footnote{This is in contrast with
vector perturbations, which cannot be generated by scalar-field perturbations
either. One can easily derive from \eqref{fullmetric} that without a 
source term the $(0i)$-component of the Einstein equation gives $\gD S_i = 0$, 
so that the vector perturbations are zero. Moreover, even if there is a 
(non-inflation) vector source at some time, one finds from the $(ij)$-component 
of the Einstein equation that vector perturbations decay as $a^{-2}$ during the 
expansion of the universe. Hence one only has to consider scalar and tensor
perturbations in inflation theory, and not vector perturbations.}
Because the scalar fields do not generate the tensor perturbations, there is 
no difference between the treatment of these perturbations in multiple-field or 
in single-field inflation. Inflation enters only by way of the background 
quantities. Hence the results derived in this subsection are not new (see 
e.g.\ \cite{Starobinsky,Mukhanovetal,MartinSchwarz} and references therein), 
but we derive them here using the methods and definitions of \cite{GNvT2}.

The $(ij)$-component of the Einstein equation gives the well-known equation of
motion for the tensor perturbation:
$h_{ij}'' + 2 \cH h_{ij}' - \gD h_{ij} = 0$.
Because $h_{ij}$ is symmetric, transverse and traceless, it has only two
independent components, which are represented by two constant polarization 
tensors $e_{ij}^A$, $A=1,2$, normalized as $e_{ij}^A e^{ij\, B} = \gd^{AB}$ and 
satisfying the same three properties as $h_{ij}$. The tensor perturbation is 
written as
$h_{ij}(\eta,\vc{x}) = \sum_{A=1}^2 (2\gk/a) \psi_A(\eta,\vc{x}) e_{ij}^A$,
where the factor $2\gk/a$ has been taken out to simplify the equation of motion 
and to obtain the correct normalization of the Lagrangean. 
After switching to Fourier modes we find the usual equation of motion for
the mode functions $\psi_{A\,\vc{k}}(\eta)$:
\equ{
\psi_{A\,\vc{k}}'' + \lh k^2 - \frac{a''}{a} \rh \psi_{A\,\vc{k}} = 0.
\labl{eqpsiA}
}
This equation is similar to equation (34) in \cite{GNvT2} for the redefined
gravitational potential $u_{\vc{k}}$, but without an inhomogeneous
multiple-field term.
An important difference between $\psi_A$ and $u$ is that the 
$\psi_A$ represent two physical degrees of freedom and can be quantized 
directly, while $u$ is not a physical degree of freedom, and had to be 
quantized indirectly by means of the scalar field degrees of freedom $q$.
There is no coupling between the two different polarizations
$A=1$ and $A=2$, as can be seen from \eqref{eqpsiA}. This means that it is 
not necessary to introduce a 2 by 2 matrix analogous to $Q$ as we had to do 
for the quantization of $q$ in \eqref{defqwithR} (as there is no coupling, 
this matrix would remain diagonal and can therefore be represented by a vector 
just as well).
The Lagrangean of the system is
$L = {\textstyle\half} (\psi_{A\, \vc{k}}')^2 - {\textstyle\half} 
( k^2 - a''/a ) (\psi_{A\,\vc{k}})^2$
and quantization is straightforward:
\equ{
\hat{h}_{ij \,\vc{k}}(\eta) = \sum_{A=1}^2 \frac{2\gk}{a} \, e_{ij}^A 
\lh \psi_{A\,\vc{k}}(\eta) \, \hat{a}^\dag_{A\,\vc{k}} 
+ \mbox{c.c.} \rh,
\labl{quanthij}
}
with the creation and annihilation operators satisfying the usual commutation 
relations. Since the different Fourier modes, as well as the
different polarizations, do not couple, we drop the subscripts $A$ and 
$\vc{k}$ for notational simplicity and consider one generic polarization 
and Fourier mode in the rest of this subsection. 

In a way analogous to (but much simpler than) the treatment for the scalar 
perturbations in subsection~3.2 of \cite{GNvT2} we can derive the 
initial conditions for $\psi$ and the canonical momentum 
$\der L/\der\psi' = \psi'$ by using the canonical commutation relations between
$\psi$ and $\psi'$ and the condition that the Hamiltonian does not contain any
particle creation or annihilation terms initially, when $k^2$ is still much
bigger than any other scale. This leads to the relations
$\psi^* \psi' - \psi {\psi'}^* = \rmi$ and $(\psi')^2 + k^2 \psi^2 = 0$,
with the solution $\psi_i = \rme^{\rmi \ga}/\sqrt{2k}$ and 
$\psi'_i = \rmi \sqrt{k/2} \, \rme^{\rmi \ga}$.
Here $\ga$ is an arbitrary phase factor, which in a way completely analogous to
the scalar case can be shown to be irrelevant to the physical
correlator, just as the whole sub-horizon region, where $\psi$ is simply
oscillating, is irrelevant. Hence we take $\ga=0$ without loss of 
generality.

Realizing that $a''/a = \cH^2 (2-\tge)$ we see that the whole treatment in
subsection~3.5 of \cite{GNvT2} of the scalar case is easily applied to this 
case as well. In a sufficiently small interval around $\eta_\cH$ (the time when 
$k=\cH$) we find to first order in slow roll with $z=k\eta$:
\equ{
\fl
\psi(z) = \sqrt{\frac{\pi}{4 k}} \sqrt{z} \, H^{(1)}_{3/2+\tge_\cH}(z)
= - \frac{\rme^{\rmi\pi\tge_\cH}}{\rmi\sqrt{2k}} 
\left [ 1 + \tge_\cH(1-\gg-\ln 2) \right ]
\lh \frac{z}{z_\cH} \rh^{-1-\tge_\cH},
\labl{psitrans}
}
where $H^{(1)}$ is a Hankel function and the expression after the last equals 
sign is only valid for $|z|\ll 1$.\footnote{Note that the change from the
complex time dependence $\rme^{\rmi k \eta}/\sqrt{2k}$ in the sub-horizon
region, satisfying the quantum commutation relation $\psi^* \psi' - \psi
{\psi'}^* = \rmi$, to the real time dependence in \eqref{psitrans} at the end
of the transition region, satisfying the classical ($\hbar \rightarrow 0$)
relation $\psi^* \psi' - \psi {\psi'}^* = 0$, is one way to see the
quantum-to-classical  transition taking place. For more information see 
\cite{PolarskiStarstates,Kieferetal}.}
On the other hand, the solution in the super-horizon region where $k \ll \cH$ is
given by
\equ{
\psi(z) = C a + D a \int^z_{z_\cH} \frac{\d z'}{a^2}.
\labl{solpsish}
}
Integrating the expression $\cH'=\cH^2(1-\tge)$, with $\tge=\tge_\cH$ taken
constant (first-order slow-roll approximation), once with respect to conformal
time we find $\cH(\eta)=-[(1-\tge_\cH)\eta]\inv$. Integrating once more, using 
$\cH = (\ln a)'$, we find the following first-order approximation for $a(z)$ 
around $\eta=\eta_\cH$:
$
a(z) = a_\cH ( z/z_\cH )^{-1-\tge_\cH}.
$
Using the identification procedure described in \cite{GNvT2} we
see that again it is the leading-order term in the expansion in $z$ of
the Hankel function in the solution for $\psi$ in the transition region that
turns into the dominant solution for $\psi$ in the super-horizon region,
i.e.\ the $C$ term in \eqref{solpsish}. From this we derive an expression for
$C$:
$C = (\sqrt{2k} \, a_\cH)\inv [ 1 + \tge_\cH(1-\gg-\ln 2) ]$.
Here we omitted some unitary factors that are irrelevant to the calculation of 
the correlator of the tensor perturbations.
As the $D$ term in \eqref{solpsish} rapidly decays compared to the $C$~term, we
do not have to determine $D$.

Using \eqref{quanthij} we then obtain the final result for~$h_{ij}$ at later
times (i.e.\ when we can neglect the $D$ term) to first order in slow roll for
super-horizon modes that crossed the horizon during slow-roll inflation:
\equ{
\hat{h}_{ij \,\vc{k}} = \frac{\sqrt{2} \, \gk}{k^{3/2}}
\, H_\cH \left [ 1 + \tge_\cH (1-\gg-\ln 2) \right ]
\sum_{A=1}^2 e_{ij}^A \, \hat{a}^\dag_{A\,\vc{k}} + \mbox{c.c.},
\labl{tensorresult}
}
where we used the identity $k=a_\cH H_\cH$. 
This result is also valid after inflation, as long as the mode $\vc{k}$ remains
super horizon. We see that this expression for $h_{ij}$ is independent of time: 
the super-horizon $h_{ij}$ is simply constant. Of course this could be
seen directly from the equation of motion for $h_{ij}$ for $k^2$ negligibly 
small (i.e.\ neglecting the $\gD h_{ij}$ term).

\section{Inflation and the CMB}
\labl{corrinfl}

In this section (plus the appendix), expressions for the scalar (adiabatic, 
isocurvature and mixing) and tensor amplitudes and spectral indices and their 
derivatives are derived. The results are discussed, paying
special attention to multiple-field effects. Moreover, a number of important 
relations between these quantities are derived, which can be used as 
observational tests to determine the presence of multiple fields during 
inflation.

\subsection{Spectral quantities from inflation}
\labl{specquantinflsec}

Having determined the relevant scalar and tensor perturbation quantities
at the time of recombination (see \eqref{Phiadafter}, \eqref{Phiisoafter} and 
\eqref{tensorresult}) we can now compute their correlators.
We define the quantities $|\gd_{\vc{k}}^X|^2$ ($X$ denoting adiabatic,
isocurvature, mixing or tensor) as
\equ{
\fl
|\gd_{\vc{k}}^{\mathrm{ad}}|^2 \equiv \frac{2 k^3}{9\pi^2} 
\langle \hat{\Phi}_{\vc{k}\,\mathrm{ad}}^2 \rangle_{t_{\mathrm{rec}}},
\qquad
|\gd_{\vc{k}}^{\mathrm{iso}}|^2 \equiv \frac{2 k^3}{9\pi^2} 
\langle \hat{\Phi}_{\vc{k}\,\mathrm{iso}}^2 \rangle_{t_{\mathrm{rec}}},
\qquad
|\gd_{\vc{k}}^{\mathrm{tens}}|^2 \equiv \frac{2 k^3}{9 \pi^2}
\langle \hat{h}^{\;}_{ij\,\vc{k}} \hat{h}^{ij}_{\vc{k}} \rangle_{t_\mathrm{rec}},
\non\\
|\gd_{\vc{k}}^{\mathrm{mix}}|^2 \equiv \frac{2 k^3}{9\pi^2} \lh
\langle \hat{\Phi}_{\vc{k}\,\mathrm{iso}} \hat{\Phi}_{\vc{k}\,\mathrm{ad}} 
\rangle_{t_{\mathrm{rec}}}
+ \langle \hat{\Phi}_{\vc{k}\,\mathrm{ad}} \hat{\Phi}_{\vc{k}\,\mathrm{iso}} 
\rangle_{t_{\mathrm{rec}}} \rh.
\labl{defamplitude}
}
Unfortunately there are different conventions in the literature regarding the
normalization factor. The normalization used here corresponds with
\cite{LiddleLyth,BunnLiddleWhite}, in which papers the relation between this
amplitude and the observations is explained extensively.
The normalization factor in, for example, \cite{MartinSchwarz}, which
was used in the table with numerical values in our paper \cite{GNvT2} as well, 
is $9/4$ times larger.

If $|\gd_{\vc{k}}^X|^2$ depends only weakly on $k$, the following
approximation can be made:
\equ{
|\gd_{\vc{k}}^X|^2 = |\gd_{\vc{k}_0}^X|^2 \lh \frac{k}{k_0} \rh^{\tn_X},
\labl{expansionamplitudeslope}
}
with $k_0$ a certain reference scale. $|\gd_{\vc{k}_0}^X|^2$ and $\tn_X$ are 
two constants, called the amplitude of the CMB power spectrum and the spectral
index, respectively. This approximation holds good over a wide range of $k$ if
$\tn_X$ is close to zero, which is the case for slow-roll inflation (see 
later). Note that originally the spectral index was defined as
$n_X = \tn_X + 1$, except for the tensor perturbations where the above 
definition was indeed used. As this was a rather unfortunate source of 
confusion, newer papers usually adopt the definition given here, but to avoid 
confusion we have added the tilde. 

With these definitions we find for the amplitudes the following expressions,
valid up to and including first order in slow roll:
\equa{
\fl
|\gd_{\vc{k}}^{\mathrm{ad}}|^2
= & \frac{\gk^2}{50 \pi^2}
\frac{H_\cH^2}{\tge_\cH} & \Bigl [ (1-2\tge_\cH)(1 + U_{P\, e}^T U_{P\, e}^{\;})
\non\\
&& \, + 2 B \lh 2\tge_\cH^{\;} + \tget^\parallel_\cH
+ 2 \tget^\perp_\cH e_2^T U_{P\, e}^{\;} + U_{P\, e}^T \gd_\cH U_{P\, e}^{\;} 
\rh \Bigr ],
\labl{CorrelatorPhi}\\
\fl
|\gd_{\vc{k}}^{\mathrm{iso}}|^2
= \;\:\, \frac{1}{36} & \frac{\gk^2}{50 \pi^2} \frac{H_\cH^2}{\tge_\cH} 
& \Bigl [ (1-2\tge_\cH) V_{e}^T V_{e}^{\;}
+ 2 B V_{e}^T \gd_\cH V_{e}^{\;} \Bigr ],
\labl{corriso}\\
\fl
|\gd_{\vc{k}}^{\mathrm{mix}}|^2
= \;\; \frac{1}{6} & \frac{\gk^2}{50 \pi^2} \frac{H_\cH^2}{\tge_\cH} 
& \Bigl [ (1-2\tge_\cH) U_{P\, e}^T V_{e}^{\;} 
+ 2 B \lh \tget^\perp_\cH e_2^T V_{e}^{\;} + U_{P\, e}^T \gd_\cH V_{e}^{\;} 
\rh \Bigr ],
\labl{corrmix}\\
\fl
|\gd_{\vc{k}}^{\mathrm{tens}}|^2 
= \frac{400}{9} & \frac{\gk^2}{50 \pi^2} \, H_\cH^2
& \Bigl [ 1 + 2 (B-1) \tge_\cH \Bigr ].
\labl{tensampl}
}
Here $U_{P\, e}$, $V_e$ and $\gd$ are defined in \eqref{defUPe} and 
\eqref{defdelta}, and $B \equiv 2 - \gg - \ln 2 \approx 0.7296$. 
Note that $U_{P\, e}^T U_{P\, e}^{\;}$ and $V_e^T V_e^{\;}$ can be of order
unity and hence the leading-order expressions include the 
$(1+U_{P\, e}^T U_{P\, e}^{\;})$, the $V_e^T V_e^{\;}$, and the 
$U_{P\, e}^T V_e^{\;}$ factors, respectively. The reader interested in more 
detail about the derivation is referred to \cite{GNvT2,thesis}.

The spectral indices $\tn_X$ can be calculated from the expressions for
$|\gd_{\vc{k}}^X|^2$ above:
\equ{
\tn_X \equiv \left. \dd[\ln |\gd_{\vc{k}}^X|^2]{\ln k} 
\right |_{\vc{k}=\vc{k}_0}
= \dd[\ln |\gd_{\vc{k}}^X|^2]{t_\cH} \dd[t_\cH]{\ln k}
= \dd[\ln |\gd_{\vc{k}}^X|^2]{t_\cH} \frac{1}{H_\cH(1-\tge_\cH)}.
\labl{defnX}
}
Here we omitted the explicit $\vc{k}=\vc{k}_0$ from the last two steps, but of
course it should be applied there as well. In the last step we used 
$\d t_\cH / \d \, \ln k = (\d \, \ln k / \d t_\cH)\inv$ and $\cH_\cH=k$.
To work out this expression we need the derivatives of $U_{P\, e}$ and $V_e$
with respect to $t_\cH$:
\equa{
\dd[U_{P\, e}^T]{t_\cH} & = \half \frac{\dot{\tge}_\cH}{\tge_\cH} \, U_{P\, e}^T
- 2 H_\cH^{\;} \tget^\perp_\cH e_2^T + \frac{\dot{a}_\cH}{a_\cH} \, U_{P\, e}^T
+ U_{P\, e}^T Q_\cH^{\;} ({Q}_\cH\inv)^{\mbox{\huge .}}
\non\\
& = H_\cH U_{P\, e}^T \lh 2 \tge_\cH^{\;} + \tget^\parallel_\cH - \gd_\cH \rh
- 2 H_\cH^{\;} \tget^\perp_\cH e_2^T,
\\
\dd[V_e^T]{t_\cH} 
& = H_\cH V_e^T \lh 2 \tge_\cH^{\;} + \tget^\parallel_\cH - \gd_\cH \rh.
}
Here we used 
$(Q_\cH\inv)^{\mbox{\Large $\cdot$}} = - Q_\cH\inv \dot{Q}_\cH^{\;} Q_\cH\inv$ 
and $\dot{Q}_\cH = H_\cH (1-\tge_\cH+\gd_\cH) Q_\cH$.
The latter result follows from the fact that $Q \propto \eta^{-\Id-\gd_\cH}$
near $\eta_\cH$ (see e.g.\ equation (63) in \cite{GNvT2}) and using the 
expression for $\cH(\eta)$ in the text below \eqref{solpsish}.
The final results to leading (first) order in slow roll are
\equa{
[\tn_{\mathrm{ad}}]_{\mathrm{l.o.}} 
= - 2 \, \frac{2 \tge_\cH^{\;} + \tget^\parallel_\cH
+ 2 \tget^\perp_\cH e_2^T U_{P\, e}^{\;} + U_{P\, e}^T \gd_\cH U_{P\, e}^{\;}}
{1 + U_{P\, e}^T U_{P\, e}^{\;}},
\labl{slopead}\\
{[\tn_{\mathrm{iso}}]_{\mathrm{l.o.}}}
= - 2 \, \frac{V_e^T \gd_\cH V_e^{\;}}{V_e^T V_e^{\;}},
\labl{slopeiso}\\
{[\tn_{\mathrm{mix}}]_{\mathrm{l.o.}}} 
= - 2 \, \frac{\tget^\perp_\cH e_2^T V_e^{\;}
+ U_{P\, e}^T \gd_\cH V_e^{\;}}{U_{P\, e}^T V_e^{\;}},
\labl{slopemix}\\
{[\tn_{\mathrm{tens}}]_{\mathrm{l.o.}}} 
= -2 \tge_\cH,
\labl{slopetens}
}
where l.o.\ means leading order. The result for the adiabatic spectral 
index was previously derived in our paper \cite{vTGNcosm}.
When comparing \eqref{CorrelatorPhi}--\eqref{tensampl} and 
\eqref{slopead}--\eqref{slopetens}, we see that we can rewrite the spectral
amplitudes to first order in slow roll in terms of the leading-order 
expressions:
\equ{
|\gd_{\vc{k}}^X|^2 = |\gd_{\vc{k}}^X|^2_{\mathrm{l.o.}}
\lh 1 + [\tn_{\mathrm{tens}}]_{\mathrm{l.o.}} 
- B [\tn_X]_{\mathrm{l.o.}} \rh.
\labl{gdkXntlo}
}
This means that using \eqref{defnX} we can immediately extend our results for 
the spectral indices to include the next-to-leading-order (i.e.\ second-order) 
terms:
\equ{
\tn_X = [\tn_X]_{\mathrm{l.o.}} \lh 1 + \tge_\cH \rh
+ \left [ \dd[\tn_{\mathrm{tens}}]{\ln k} \right ]_{\mathrm{l.o.}}
- B \left [ \dd[\tn_X]{\ln k} \right ]_{\mathrm{l.o.}}.
\labl{tnXntlo}
}
The $\d \tn_X / \d \, \ln k$ are given in the appendix.

\subsection{Discussion}
\labl{discusssec}

We continue with a discussion of the results derived in the previous subsection, 
leading to several important conclusions. Some of these were already given in 
\cite{GNvT2}, but are repeated here for completeness' sake.

We see that the $\tn_X$ contain only slow-roll terms, and thus the
$|\gd_{\vc{k}}^X|^2$ do indeed depend on $k$ very weakly. In other words, the
spectrum predicted by slow-roll inflation is nearly scale-invariant. Of course 
if $V_e=0$, there are no isocurvature perturbations and $\tn_{\mathrm{iso}}$ and
$\tn_{\mathrm{mix}}$ become meaningless. The derivatives of the spectral indices
$\d \tn_X / \d \, \ln k$ are second order in slow roll. This might be too small
according to recent WMAP data \cite{WMAP,WMAPinfl}, where a 
$\d \tn_X / \d \, \ln k$
of the same order of magnitude as $\tn_X$ is claimed, leading to a problem for
slow-roll models of inflation in general. However, as explained in
\cite{WMAPparam}, this result is only obtained when combining WMAP with
other data sets, especially \mbox{Lyman-$\ga$} data. Hence this result is as yet
controversial, with several authors \cite{Bargeretal,Bridleetal,Kinneyetal,
LeachLiddle} claiming that a constant $\tn_X$ is perfectly consistent with 
the data.

The explicit multiple-field terms in the amplitudes and the spectral indices 
are the contributions of the terms $U_{P\, e}$ and $V_e$, which are absent in 
the single-field case (setting them equal to zero we obtain the well-known
single-field results, see e.g.\ \cite{MartinSchwarz}). 
There are no isocurvature and mixing contributions in the
single-field case. Since both $U_{P\, e}$ and~$V_e$ are to a large extent 
determined by $\tget^\perp$ (see their definitions in \eqref{defUPe}), 
we can draw the important conclusion that the behaviour of 
$\tget^\perp$ during the last 60 e-folds of inflation is crucial in order to 
determine whether multiple-field effects are important. 
On the other hand, the fact that $U_{P\, e}$ depends on $\tget^\perp$ should not
be taken as an indication that it is a first-order quantity: as was shown in the
example in \cite{GNvT2} $U_{P\, e}$ can be of zeroth-order importance (because
of the integration interval).

The fact that the entropy perturbations act as sources for the adiabatic
perturbation (the $U_{P\, e}$ terms) naturally leads to correlations between 
adiabatic and isocurvature perturbations (described by the mixing amplitude), as
had been realized before (see e.g.\ \cite{Langlois,Gordonetal,Bartoloetal1}).
Note that even if $U_{P\, e}=0$ there is still one other term in the mixing 
amplitude, although merely of first order in slow roll. Only if 
$\tget^\perp_\cH$ vanishes as well, the correlations are completely absent. 
However, at least in the context of slow roll the situation where $U_{P\, e}=0$ 
while $\tget^\perp_\cH,V_e \neq 0$ is not possible, because the $e_2^T Q/a$ 
under the integral in the definition of $U_{P\, e}$ cannot change sign. 
Anyhow, if $\tget^\perp=0$ everywhere during the last 60 e-folds, there are 
certainly no correlations. (The authors of \cite{Gordonetal} studied
the two-field case and found the derivative of the angle that
parametrizes the influence of the second field on the background trajectory to
be the relevant parameter. In the two-field limit this parameter corresponds
with $\tget^\perp$, but our result is valid for an arbitrary number of fields.
A general discussion of two-field models has recently been given in
\cite{Tsujikawaetal}.)

The gravitational potential only depends on the {\it total} entropy 
perturbation~$\tS$, independently of the total number of fields and the actual 
number of independent entropy perturbations. In our basis $\tS$ depends 
directly on the $e_2$ component of $q$ only (see the expression for $V_e$ in 
\eqref{defUPe} or \eqref{defVe2}). (Of course the other components of $q$
influence the equation of motion for $q_2$ and cannot be neglected in general.)
This is the main reason why we only consider this total entropy perturbation 
and a total isocurvature amplitude in this paper. However, for a complete 
understanding of the CMB power spectrum it will probably be necessary to 
investigate the individual entropy perturbations as well.\footnote{For example,
one can prove that if there is not only baryonic matter, radiation and cold 
dark matter, but hot dark matter and/or quintessence as well, their individual
entropy perturbations enter the expression for the Sachs-Wolfe effect (see
\cite{thesis}). An investigation of the individual entropy 
perturbations in a model with photons, baryons, neutrinos, and cold dark matter 
was given in \cite{Bucheretal}. The same authors investigated the detectibility 
of these entropy perturbations in upcoming experiments in \cite{Bucheretal2} 
and concluded that we will have to wait for the Planck satellite \cite{Planck} 
for sufficient discriminating power.}

Using the concept of slow roll for the perturbations the quantity~$U_{P\, e}$ 
can be rewritten in terms of background quantities only, as was discussed in
\cite{GNvT2} (see equation~(69) in that paper).
Because slow roll was used in the derivation, that expression is
in principle not valid at the very end of inflation. (Note that until this 
point slow roll was only used during a small transition region around the 
time of horizon crossing deep within the inflationary era.) If it does
indeed give a bad approximation for $U_{P\, e}$, for example if $\tget^\perp$
grows very large, a more careful treatment of the transition at the end of
inflation is necessary. However, in other cases the contribution to the integral
near the end of inflation can be negligible, for example if $\tget^\perp$ goes
sufficiently rapidly to zero. In those cases the details at the end of inflation
are unimportant for the adiabatic amplitude \eqref{CorrelatorPhi}. 
An example of this latter case was discussed in \cite{GNvT2}.
Unfortunately $V_e$ depends very much on the details of the transition at the 
end of inflation, so that for an accurate calculation a model of this 
transition has to be assumed.
It also depends on the behaviour of $\tS$ during radiation and matter
domination, as shown in the derivation of \eqref{defVe2}, but this aspect is
investigated in section~\ref{entropypertsec}.
However, one can still draw the conclusion that, if $\tget^\perp$ goes to zero 
at the end of inflation, the isocurvature perturbations are expected to be
negligible compared with the adiabatic one (neglecting possible
amplifying mechanisms during the transition and preheating).

Compared with the scalar amplitudes, an overall factor of $1/\tge_\cH$ is 
missing in the expression for the tensor amplitude, showing that the tensor 
contribution to the CMB from slow-roll inflation will generically be smaller 
than the scalar one. 
Moreover, this means that the tensor amplitude depends only on $H_\cH^2$ (to 
leading order), thus allowing a direct determination of the inflationary energy 
scale. Another important difference between the tensor and scalar quantities is 
that the tensor ones do not depend on multiple-field terms. As explained in
subsection~\ref{tensorpert} this is because the tensor perturbations are not 
generated by the scalar-field perturbations and hence do not see the difference 
between multiple- and single-field inflation.
It is this fact that allows us to derive, in the following subsection, an 
important relation that can be used as an observational test for the importance 
of multiple-field effects.

\subsection{Observational tests and consistency relations} 
\labl{obstestsec}

From the expressions for the spectral amplitudes and indices
(\eqref{CorrelatorPhi}--\eqref{tensampl} and \eqref{slopead}--\eqref{slopetens}) 
we can derive some important relations between these quantities. Note that we 
only use the general expressions for these amplitudes and indices in the 
derivations, not the specific form of, for example, $V_e$. Therefore the 
results are valid very generally, and do not depend on the details of the 
evolution of the entropy perturbations after inflation (as was explained in 
the derivation of \eqref{defVe2}, these only change the definition of $V_e$).

We define the tensor to scalar ratio $r$ as 
$r \equiv |\gd_{\vc{k}}^{\mathrm{tens}}|^2 / |\gd_{\vc{k}}^{\mathrm{ad}}|^2$.
Then we find the following relation to leading order:
\equ{
\fl
r = \frac{400}{9} \frac{\tge_\cH}{1+U_{P\, e}^T U_{P\, e}^{\;}}
= - \frac{200}{9} \frac{\tn_{\mathrm{tens}}}{1+U_{P\, e}^T U_{P\, e}^{\;}}
\quad\Rightarrow\quad
U_{P\, e}^T U_{P\, e}^{\;} = -1 - \frac{200}{9} \frac{\tn_{\mathrm{tens}}}{r}.
\labl{consistrel}
}
This expression is a very important result: it gives a 
relation between observable quantities and the length of the vector $U_{P\, e}$ 
that encodes the effects of multiple fields.  
Once observations have become good enough to determine the tensor to scalar
ratio $r$ and the tensor spectral index $\tn_{\mathrm{tens}}$ independently, 
this relation offers an observational test to check whether multiple-field 
effects are important or not.

Using \eqref{gdkXntlo} and \eqref{tnXntlo} we can extend this result to 
next-to-leading order:
\equ{
\fl
U_{P\, e}^T U_{P\, e}^{\;} = -1 - \frac{200}{9} \frac{\tn_{\mathrm{tens}}}{r}
\left [  1 - (B-{\textstyle \half}) \tn_{\mathrm{tens}} + B \, \tn_{\mathrm{ad}} 
+ (B-1) \frac{1}{\tn_{\mathrm{tens}}} \dd[\tn_{\mathrm{tens}}]{\ln k} \right ].
\labl{consistrel2}
}
This does not change the above conclusion: it is still a relation that allows
the multiple-field quantity $U_{P\, e}^T U_{P\, e}^{\;}$ to be determined from
observations.\footnote{There is one other 
effect that might lead to corrections to this formula: non-vacuum initial 
states caused by trans-Planckian physics \cite{HuiKinney}. However, this 
effect is expected to be small: deviations from the vacuum initial state should 
not be too large, otherwise the particle background dominates over the potential 
energy of the inflaton and there is no (standard) inflation \cite{Tanaka}. For a
recent review of trans-Planckian effects see \cite{MartinBrandenberger}.} 
We can also (in principle) determine the other multiple-field quantities 
$V_{e}^T V_{e}^{\;}$ and $U_{P\, e}^T V_{e}^{\;}$ from the observations:
\equ{
V_{e}^T V_{e}^{\;} = - 800 \, \frac{\tn_{\mathrm{tens}}}
{|\gd_{\vc{k}}^{\mathrm{tens}}|^2 / |\gd_{\vc{k}}^{\mathrm{iso}}|^2},
\qquad
U_{P\, e}^T V_{e}^{\;} = - \frac{400}{3} \frac{\tn_{\mathrm{tens}}}
{|\gd_{\vc{k}}^{\mathrm{tens}}|^2 / |\gd_{\vc{k}}^{\mathrm{mix}}|^2},
}
with similar extensions to the next order as in \eqref{consistrel2}.

Next to the above results valid for an arbitrary number of fields, we can find
some further results in the case of one or two fields only. These are the
so-called consistency relations, which are based on the fact that in those two
cases we have more observational quantities than inflationary parameters.
In the single-field case the unknowns are $H_\cH$, $\tge_\cH$ and
$\tget^\parallel_\cH$, while the scalar and tensor amplitudes and indices give 
four observational quantities. Hence we can express one of these in terms of 
the others: a consistency relation. Conventionally this is written as follows
(valid to leading order) \cite{LythLiddle,LiddleLyth,Lidseyetal}:
\equ{
r = - \frac{200}{9} \, \tn_{\mathrm{tens}}
\qquad\qquad\qquad\;\mbox{(one field),}
\labl{echteconsistrel1f}
}
a relation that follows immediately from the single-field limit of
\eqref{consistrel}.

In the two-field case, there are four more inflationary parameters:
$\tget^\perp_\cH$, $U_{P\, e}$ and $V_e$ (both are vectors, but in the two-field
case there is only one non-zero component), and $\gd_\cH$ (this is a matrix, but
in the two-field case there is only one unknown component). As there are also
four more observational (in principle at least) quantities, there is still one
consistency relation, which can be written as (again only valid to leading
order):
\equ{
r = - \frac{200}{9} \, \tn_{\mathrm{tens}} \lh 1 - r_{\mathrm{mix}}^2 \rh
\qquad\mbox{(two fields).}
\labl{echteconsistrel}
}
Here $r_{\mathrm{mix}}^2$ is defined as
\equ{
r_{\mathrm{mix}}^2 \equiv \frac{|\gd_{\vc{k}}^{\mathrm{mix}}|^4} 
{|\gd_{\vc{k}}^{\mathrm{ad}}|^2 |\gd_{\vc{k}}^{\mathrm{iso}}|^2}
= \frac{(U_{P\, e}^T V_{e}^{\;})^2}
{(1 + U_{P\, e}^T U_{P\, e}^{\;}) V_{e}^T V_{e}^{\;}},
\labl{defrmix}
}
where the second expression is valid to leading order.
In the two-field case where the vectors $U_{P\, e}$ and $V_e$ have only one
non-zero component (denoted by $U$ and $V$, respectively), the $V$ drops out 
and one has $r_{\mathrm{mix}}^2 = U^2/(1+U^2)$, which leads to the expression in
\eqref{echteconsistrel}.
This two-field consistency relation was also derived in a different way in
\cite{Bartoloetal,Wandsetal2}.\footnote{In \cite{Bartoloetal} a second 
consistency relation for the two-field case was derived as well. However, this 
relation is not valid in general, as was also pointed out in \cite{Wandsetal2}, 
but only if the inflation model satisfies certain specific additional 
conditions (namely that the (22)-component of the matrix $\gd_\cH$ is equal to
$2\tge_\cH+\tget^\parallel_\cH$).}
Of course the two-field consistency relation reduces to the single-field one in
the appropriate limit because $r_{\mathrm{mix}}$ is zero then.
In the case of three or more fields the number of inflationary parameters
increases (with four in the case of three fields: one additional component in
$U_{P\, e}$ and $V_e$ and two in $\gd_\cH$) without an increase in the number of
observational quantities (at least when considering only the total isocurvature
perturbation). Hence there are no consistency relations in those cases.
For some consistency relations concerning the $\d \tn_X / \d \, \ln k$ see the
appendix.

To conclude this section: assuming that we will be able to measure the scalar 
and tensor amplitudes and indices with sufficient accuracy in the near future
(which of course implicitly assumes that they are large enough to be measured), 
we can then first use equation \eqref{consistrel} (or \eqref{consistrel2}) to 
check if multiple-field effects are significant at all. Next we can use the 
consistency relation \eqref{echteconsistrel} to distinguish between the cases of
two or more fields. Note that an accurate measurement of the isocurvature and
mixing quantities is only necessary for the second step, not for the first. An
accurate measurement of the tensor amplitude and spectral index is, however,
essential for both steps (for a recent study on detectability issues see 
\cite{SongKnox}).

\section{Entropy perturbations}
\labl{entropypertsec}

In the derivation of the isocurvature perturbation \eqref{Phiisoafter}, it was
first assumed that the total entropy perturbation $\tS$ is constant during 
radiation and matter domination on super-horizon scales, leading to equation
\eqref{defUPe} for $V_e$. We then considered the possibility that $\tS$ is not
constant, finding that \eqref{Phiisoafter} is still correct, but now with a more
general expression \eqref{defVe2} for $V_e$. In this section, we investigate the 
total entropy perturbation during radiation and matter domination to determine 
its evolution with time.

\subsection{Total entropy perturbation}

We consider a universe filled with an arbitrary number $N$ of energy components, 
labelled by the subscript $i$. The different components each have a 
pressure~$p_i$ and an energy density $\gr_i$, as well as pressure and density 
perturbations.
For the total pressure $p = \sum_i p_i$ and the total energy density 
$\gr = \sum_i \gr_i$ we have
\equ{
\fl
w \equiv \frac{p}{\gr},
\qquad
c_s^2 \equiv \frac{\dot{p}}{\dot{\gr}},
\qquad
\dot{\gr} + 3 H \gr (1+w) = 0,
\qquad
\dot{w} = -3 H (c_s^2 - w)(1+w).
\labl{defw}
}
The first two expressions are definitions ($w$ is called the equation of state
parameter and $c_s^2$ the sound velocity), the third one is the energy-momentum
conservation condition $D_\mu T^\mu_{\;\:0}=0$, and the last equation follows by
writing $\dot{w} = (\dot{p}/\dot{\gr} - p/\gr)\dot{\gr}/\gr$.
For the individual components we define analogous quantities: 
$w_i \equiv p_i/\gr_i$ and $c_i^2 \equiv \dot{p}_i/\dot{\gr}_i$
(note that in general $w \neq \sum_i w_i$ and $c_s^2 \neq \sum_i c_i^2$)
and find
\equ{
\fl
\dot{\gr}_i + 3 H \gr_i (1+w_i) = 3 H C_i,
\qquad
\dot{w}_i = -3 H (c_i^2-w_i) \left [ (1+w_i) - \frac{C_i}{\gr_i} \right ],
\labl{dotwi}
}
with $C_i$ a measure of the interactions between the different components
satisfying $\sum_i C_i = 0$.
In the following we make two assumptions regarding the separate components,
which for the rest are completely arbitrary:
\begin{enumerate}
\item All components behave as ideal fluids with a constant $w_i$.
\item There are no interactions: $C_i=0$ for all $i$.
\end{enumerate}
From equation \eqref{dotwi} we see that this automatically means that 
$c_i^2 = w_i$ (the square in $c_i^2$ is just convention; $c_i^2$ can be
negative). Moreover, a constant $w_i$ also means that 
$\gd p_i / \gd\gr_i = w_i$. 

Let us remark briefly on the assumption of no interactions with regard to a 
real model. One can think of the following situation. Of the multiple 
scalar fields during inflation, one has decayed to all the Standard Model 
particles, while the other fields have decayed to various kinds of dark 
matter. Then there are by construction no entropy perturbations between the
Standard Model components on super-horizon scales and the absence 
or presence of interactions here is irrelevant. There are only entropy
perturbations between the dark matter and Standard Model components and between
the dark matter components themselves, where the assumption of no interactions 
seems quite plausible.
It is usually assumed (see e.g.\ \cite{PolarskiStar,Langlois}) that including
interactions will have the effect of wiping out the isocurvature perturbations.
However, it will be interesting to check this more carefully in the near 
future.

To rewrite $\tS$ (defined as 
$\tS \equiv \frac{1}{4}(\gd p-c_s^2 \gd\gr)/(p-c_s^2 \gr)$) we need some 
auxiliary results. 
In the first place we have that $p = \sum_i w_i \gr_i$, so that
\equ{
c_s^2 = \frac{\dot{p}}{\dot{\gr}} = \sum_i \frac{\gr_i(1+w_i)}{\gr(1+w)}\, w_i,
}
where we used \eqref{dotwi} and \eqref{defw} for $\dot{\gr}_i$ and $\dot{\gr}$.
Using this result the numerator of $\tS$ is rewritten as follows:
\equa{
\gd p - c_s^2 \gd\gr & = \sum_k (w_k - c_s^2) \gd\gr_k
\non\\
& = \frac{1}{\gr(1+w)} \sum_{k,l} \gr_l (1+w_l)(w_k-w_l) \gd\gr_k
\non\\
& = \frac{1}{\gr(1+w)} \half \sum_{k,l} \gr_k \gr_l (1+w_k)(1+w_l)(w_k-w_l) 
S_{kl}
\labl{gdpmincs2gdgr}
}
with
\equ{
S_{kl} \equiv \frac{\gd\gr_k}{\gr_k(1+w_k)} - \frac{\gd\gr_l}{\gr_l(1+w_l)}.
\labl{defSkl}
}
In the last step of \eqref{gdpmincs2gdgr} we symmetrized the expression in $k$ 
and $l$. Completely analogously we find for the denominator
\equ{
p - c_s^2 \gr = \sum_k (w_k - c_s^2) \gr_k
= - \frac{1}{\gr(1+w)} \half \sum_{k,l} \gr_k \gr_l (w_k-w_l)^2.
\labl{pmincs2gr}
}
Our final result for the total entropy perturbation is then
\equ{
\tS = - \frac{1}{4} \frac{\sum_{k,l} \gr_k \gr_l (1+w_k)(1+w_l)(w_k-w_l) S_{kl}}
{\sum_{k,l} \gr_k \gr_l (w_k-w_l)^2}.
\labl{Stildecomp}
}

The $S_{kl}$ are the individual entropy (or isocurvature) perturbations
\cite{KodamaSasaki,HwangNoh,Gordonetal} (see also \cite{Maliketal}). 
They are antisymmetric in $k$ and $l$, and in addition one has 
$S_{kl} = S_{km}-S_{lm}$. This means that the matrix $S_{kl}$ contains $(N-1)$ 
independent elements.
One can take a single reference component~$0$ and define 
$S_k \equiv S_{k0}$, so that $S_{kl} = S_k - S_l$, with of course $S_0=0$.
Hence if we have a system of $N$ components, there are in general
$1$ adiabatic and $(N-1)$ entropy perturbations. 
The combination $\tS$ of these $(N-1)$ entropy perturbations that enters as the 
source term into the equation for $\Phi$ is what we call the total entropy 
perturbation. 
In the case of inflation with $N$ fields, the adiabatic
perturbation corresponds with the $e_1$ direction in our basis, while the
perturbations in the $(N-1)$ other directions are isocurvature perturbations. 
The total entropy perturbation $\tS$ then corresponds exactly with the $e_2$ 
direction in our basis, see \eqref{tSinfl}.
The entropy perturbations $S_{kl}$ are gauge-invariant by definition in the 
absence of interactions, see e.g.\ \cite{HwangNoh}, and can even be defined in
such a way that they are gauge-invariant when interactions are included, see
\cite{Maliketal}. The total entropy perturbation $\tS$ defined in 
\eqref{defspgr} is always gauge-invariant.

Working out \eqref{Stildecomp} in the case of an arbitrary number of matter 
components labelled by $m$ (e.g.\ baryons or cold dark matter with $w_m=0$) and 
an arbitrary number of radiation components labelled by $r$ (e.g.\ photons or 
hot dark matter with $w_r=1/3$), we obtain
\equ{
\tS = \frac{\sum_{m,r} \gr_m \gr_r S_{mr}}{\sum_{m,r} \gr_m \gr_r}
= \frac{\sum_m \gr_m S_m}{\sum_m \gr_m}
- \frac{\sum_r \gr_r S_r}{\sum_r \gr_r}.
\labl{tSmatrad}
}
In the last step we have singled out one of the radiation components, for
example the photons $\gg$, as the reference component, so that 
$S_m = S_{m\gg}$ and $S_r = S_{r\gg}$. 
(One could just as well choose one of the matter components as reference. The 
only difference with \eqref{tSmatrad} is then an overall minus sign.) 
In the case of a simple two-component system consisting of photons $\gg$ and 
one cold dark matter component $C$, which is the case usually considered in 
inflationary literature (see e.g.\ \cite{PolarskiStar,Langlois,Gordonetal,
Bartoloetal}), this result simplifies to
$\tS = S_{C} =  (\gd\gr_{C}/\gr_{C}) - \frac{3}{4} (\gd\gr_\gg/\gr_\gg)$.

\subsection{Time dependence of $\tS$}

Next we derive an expression for the time derivative of $\tS$. 
First we need an equation of motion for $\gd\gr_i$. This equation can be derived
by working out the condition $D_\mu T^\mu_{\;\: 0}=0$ to first order in the
perturbations. However, as we are only interested in the super-horizon modes, it
is simpler to use the method of varying the background equation. (A description
of this method can be found in e.g.\ subsection~3.4 of \cite{GNvT2} and 
references therein.) From \eqref{dotwi}, with the two assumptions of ideal 
fluids without interactions, we then find
\equ{
\gd\dot{\gr}_i + 3 H \gd\gr_i (1+w_i) - 3 \dot{\Phi} \gr_i (1+w_i) = 0,
\labl{dergdgri}
}
using $\gd H=(\gd\ln a)^{\mbox{\Large $\cdot$}}=-\dot{\Phi}$. Equation
\eqref{dergdgri} can also be found in \cite{HwangNoh}. From this result 
together with \eqref{dotwi} we easily derive that 
\equ{
\dot{S}_{kl} = \frac{\gd\gr_k}{\gr_k(1+w_k)} 
\lh \frac{\gd\dot{\gr}_k}{\gd\gr_k} - \frac{\dot{\gr}_k}{\gr_k} \rh
- ( k \leftrightarrow l ) = 0.
}
When differentiating $\tS$, given in \eqref{Stildecomp}, this means that the
time dependence is completely determined by the background quantities. 
We find after a long calculation
\equa{
\fl
\dot{\tS} & = \frac{3}{4} H \frac{\sum_{i,j,k,l} [ (w_k+w_l)-(w_i+w_j) ]
(w_k-w_l) (w_i-w_j)^2 (1+w_k)(1+w_l) \gr_i \gr_j \gr_k \gr_l S_{kl}}
{\lh \sum_{i,j} (w_i-w_j)^2 \gr_i \gr_j \rh^2}
\non\\
\fl
& = \frac{3}{4} H \frac{\sum_{i,j,k} w_i(w_i-w_j)[w_k^2 - w_i w_k + w_i w_j]
\gr_i \gr_j \gd\gr_k}{(1+w) (w-c_s^2)^2 \gr^3}.
\labl{tSder}
}
The second form of the result comes about after inserting the definition of
$S_{kl}$ \eqref{defSkl}, substantial index manipulation in the numerator and
using \eqref{pmincs2gr} in the denominator. It is more compact, but in some
ways the first expression is more useful. From the first expression we can see
immediately that the time derivative of $\tS$ will be zero in the case of 
components with only two different values of $w_i$ (one cannot make all three 
of $[(w_k+w_l)-(w_i+w_j)]$,  $(w_i-w_j)$ and $(w_k-w_l)$ unequal to zero in that
case). Hence we can draw the important conclusion that in a universe consisting
only of matter and radiation components, $\tS$ remains constant on super-horizon
scales, irrespective of how  many kinds of matter and radiation there are
(provided that the two assumptions  of constant $w_i$ and no interactions are
valid). This includes a universe with an arbitrary  number of hot and cold dark
matter components. 

Let us also give the result for 
$\dot{\tS}$ in the case that we relax the second assumption, that is, if we 
include interactions. The sound velocity and total entropy perturbation are 
then given by
\equ{
\fl
c_s^2 = \sum_i \frac{\gr_i(1+w_i)-C_i}{\gr(1+w)}\, w_i,
\qquad
\tS = \frac{1}{4} \frac{\sum_{k,l} (1+w_l) (w_k-w_l) \gr_l \gd\gr_k 
+ w_l C_l \gd\gr_k}
{\sum_{k,l} (1+w_l) (w_k-w_l) \gr_k \gr_l + w_l C_l \gr_k}.
} 
(Since $\dot{S}_{kl}$ is no longer zero, it is not useful to rewrite $\tS$ in 
terms of the individual entropy perturbations in this case.) 
After a long calculation with substantial index manipulation, during which one 
should keep in mind that $\sum_i C_i = 0$, we finally obtain:
\equa{
\fl
\dot{\tS} = \frac{3}{4} H \left [ (1+w) (w-c_s^2)^2 \gr^3 \right ]\inv
\sum_{i,j,k} \Bigl \{
w_i(w_i-w_j)(w_k^2-w_i w_k+w_i w_j) \gr_i \gr_j \gd\gr_k 
\non\\
+ w_j(w_i-w_k)(1+2w_i-w_j+w_k) \gr_i C_j \gd\gr_k
- w_i w_j C_i C_j \gd\gr_k 
\non\\
+ w_j(w_i-w_k) \gr_i \frac{\dot{C}_j}{3 H} \gd\gr_k
- w_i w_k (w_i-w_j) \gr_i \gr_j \gd C_k + w_j w_k \gr_i C_j \gd C_k
\Bigr \}.
}
A further treatment of the interacting case is postponed to a future
publication.

Next we consider what happens in the case that there are components with three
different values of $w_i$, for example matter, radiation and a cosmological 
constant, again neglecting interactions. With more than two different
components $\tS$ is in general not constant. Even though  \eqref{tSder} gives
an expression for its time derivative, it is in general  difficult to solve
explicitly because the time dependence of the perturbations  $\gd\gr_k$ is not
trivial to determine: from \eqref{dergdgri} we see that it  depends on
$\dot{\Phi}$ so that the equations for $\tS$ and $\Phi$ have to be  solved
together. However, it turns out that in the special case of components with 
three different values of $w_i$, the time dependence of $\tS$ is determined by 
the background energy densities only and does not depend on $\Phi$.

The main idea is that we take another time derivative and calculate 
$\ddot{\tS}$. For three components we find that the numerator of $\dot{\tS}$ 
in \eqref{tSder} can be written as 
\equ{
\fl
g(w_1,w_2,w_3) \gr_1 \gr_2 \gd\gr_3 + g(w_3,w_1,w_2) \gr_1 \gr_3 \gd\gr_2 
+ g(w_2,w_3,w_1) \gr_2 \gr_3 \gd\gr_1, 
\labl{numerator3}
}
with $g(a,b,c) = (a-b)^2 (c^2 + a b - c(a+b))$. 
Taking the time derivative of this expression using \eqref{dotwi} and 
\eqref{dergdgri} we find simply $-3 H(3+w_1+w_2+w_3)$ times the same 
expression \eqref{numerator3} (the $\dot{\Phi}$ terms from \eqref{dergdgri} 
exactly cancel, which can even be proved to be true for an arbitrary number of 
components). Hence we conclude that in the three-component case 
$\ddot{\tS} \propto \dot{\tS}$, where the proportionality factor only depends 
on the background energy densities, not on the energy density perturbations. 
This means that, given initial conditions for $\tS$ and $\dot{\tS}$, one can
explicitly solve for $\tS(t)$, without needing solutions for $\gd\gr_i(t)$ and
$\Phi(t)$. For four or more components this is no longer true.

As an explicit example we consider a system with matter $m$, radiation $r$ and 
a cosmological constant $\gL$, where $w_m=0$, $w_r=1/3$, and $w_\gL=-1$. 
Then we have
\equ{
\tS_{,NN} = 4 \, \tS_{,N} \,
\frac{\gr_m^2 \gr_r + \gr_m \gr_r^2 - 9 \gr_m^2 \gr_\gL
- 16 \gr_r^2 \gr_\gL - 23 \gr_m \gr_r \gr_\gL}
{(\gr_m+{\textstyle\frac{4}{3}}\gr_r)
(\gr_m \gr_r + 9 \gr_m \gr_\gL + 16 \gr_r \gr_\gL)}.
\labl{tSNN}
}
Even though this is not during inflation, it turns out to be convenient to use
the number of e-folds $N$ as time variable to remove the $H$ and its
derivative. We choose $N$ to be zero at the present time and negative
before that. 
The functions $\gr_i(N)$ can be determined from \eqref{dotwi}: 
$\gr_i(N) = \gO_i \gr_c \exp(-3(1+w_i)N)$, where $\gO_i \equiv \gr_i(0)/\gr_c$ 
is the present density parameter of component $i$ and 
$\gr_c \equiv 3 H_0^2/\gk^2$ is the present critical density, which drops 
out of the equations, however.
Using data from WMAP \cite{WMAP} we have
\equ{
\fl
\gO_m = 0.3, \qquad \gO_r = 5 \cdot 10^{-5}, \qquad \gO_\gL = 0.7,
\qquad N_{\mathrm{eq}} = -8.7, \qquad N_{\mathrm{rec}} = -7.0,
}
where the subscript `eq' denotes matter-radiation equality. 

Equation \eqref{tSNN} can easily be solved numerically. However, using the
approximation that $\gr_r \mg \gr_m \mg \gr_\gL$ for $N_* \leq N \leq
N_{\mathrm{eq}}$ and $\gr_m \mg \gr_r \mg \gr_\gL$ for 
$N_{\mathrm{eq}} \leq N \leq N_{\mathrm{rec}}$ we can also find an analytical
solution that agrees very well with the exact numerical one. The result is
\equ{
\fl
\tS(N) = \cases{\tS^* + {\textstyle \frac{1}{3}} \tS_{,N}^* \lh \rme^{3(N-N_*)} 
- 1 \rh
& for $N_* \leq N \leq N_{\mathrm{eq}}$,\\
\tS^* - {\textstyle \frac{1}{3}} \tS_{,N}^* + {\textstyle \frac{1}{4}} 
\tS_{,N}^* \rme^{3(N_{\mathrm{eq}}-N_*)} \lh {\textstyle \frac{1}{3}} 
+ \rme^{4(N-N_{\mathrm{eq}})} \rh
& for $N_{\mathrm{eq}} \leq N \leq N_{\mathrm{rec}}$,}
}
where the superscript $*$ denotes evaluation at the beginning $N_* \sim -60$ 
of the radiation-dominated era.
To determine if the time dependence of $\tS$ leads to significant effects we
must have an estimate for $\tS_{,N}^*$. Evaluating \eqref{tSder} and
\eqref{Stildecomp} at $N_*$ (where $\gr_r \mg \gr_m \mg \gr_\gL$) for the 
situation under consideration and combining them we find
\equ{
\tS_{,N}^* = -48 \, \frac{\gO_\gL}{\gO_m} \, \rme^{3 N_*} \, \tS^*.
}
Here we have taken $\gd\gr_\gL^*=0$, assuming $\gL$ to be a pure cosmological
constant. This means that
\equ{
\tS(N_{\mathrm{rec}}) = \tS^* - 12 \, \frac{\gO_\gL}{\gO_m} \,
\rme^{4 N_{\mathrm{rec}}-N_{\mathrm{eq}}} \, \tS^*
= \lh 1 - 10^{-7} \rh \tS^*.
}
Hence the effect of the fact that $\tS$ is not constant in this system is
completely negligible. This system is practically equivalent to one without a
cosmological constant, which was of course to be expected as the energy density
of the cosmological constant is so much smaller than that of matter and
radiation before recombination.
(As soon as one takes a third component with $\gd\gr_3^*/\gr_3^* \mg \tS^*$ or
with $w_3 \neq -1$, it is no longer possible to express $\tS_{,N}^*$ in terms 
of $\tS^*$ only. Even though the solution for $\tS(N)$ can still be calculated
explicitly in these cases, it becomes more difficult to determine the relative
importance of the time dependence of $\tS$.)

\section{Summary and conclusions}
\labl{obsconcl}

In this paper, we investigated some consequences of multiple-field inflation for 
the cosmic microwave background radiation. Building on the theory of 
\cite{GNvT2} we derived expressions for the amplitudes 
\eqref{CorrelatorPhi}--\eqref{tensampl} and the spectral indices
\eqref{slopead}--\eqref{slopetens}, \eqref{tnXntlo} of the CMB power spectrum, 
all valid to next-to-leading order in slow roll. We also derived expressions
valid to leading order for the derivatives of the spectral indices
\eqref{dnad}--\eqref{dntens}  in the appendix. All this in the context of a
very general inflation theory with  an arbitrary number of real scalar fields
that may be the coordinates of a non-trivial field manifold (i.e.\ have
non-minimal kinetic terms). There are four different versions of all these
spectral quantities: three related to the scalar perturbations (adiabatic,
(total) isocurvature and  mixing between those two) and one related to the
tensor perturbations.

These expressions were discussed in subsection~\ref{discusssec}. To summarize, 
multiple-field effects can be important for the scalar spectral quantities, not 
only for the isocurvature and mixing components (which are absent in the 
single-field case), but also for the adiabatic ones. 
In all multiple-field terms the slow-roll function $\tget^\perp$, which measures
the size of the acceleration perpendicular to the field velocity, plays 
a key role; if it is negligible during the last 60 e-folds of 
inflation, multiple-field effects are unimportant. Unfortunately, to work out
the expressions for the multiple-field terms explicitly, especially for the
isocurvature perturbations, a careful analysis of the transition at the end of 
inflation as well as of the era of (p)reheating is in general required.
However, as was shown in \cite{GNvT2}, there is a wide class of models where 
$\tget^\perp$ goes to zero at the end of inflation (while being non-negligible
before that). Then the integral expression for the multiple-field contributions 
to the adiabatic perturbation can be worked out explicitly without knowing the 
details of the transition, while isocurvature perturbations are expected to be 
unimportant in this case (barring possible amplification mechanisms during 
preheating). Even in those models multiple-field effects in the adiabatic 
component can be of leading-order importance.
The leading-order terms of the spectral indices are of first order in slow roll,
while their derivatives are of second order. Hence multiple-field slow-roll
inflation generically predicts a CMB power spectrum that is close to 
scale-invariant. The tensor spectral quantities do not depend on multiple-field
effects.

From the expressions for the amplitudes and spectral indices we derived some
important relations, which are valid very generally and do not depend on the
details of what happens at the end of and after inflation. Most important is 
equation \eqref{consistrel} (or its extension to next-to-leading order 
\eqref{consistrel2}), which gives the size of the multiple-field contribution to
the adiabatic perturbation in terms of the (in principle) observable
spectral quantities. In other words, this relation provides an observational
test to determine if multiple-field effects play a significant role during the
last 60 e-folds of inflation. It does require, however, a sufficiently accurate
measurement of the tensor spectral quantities.

In the case of only one or two fields, we have the special situation that there
is one more observational quantity than there are inflationary parameters.
This means that we can then derive a consistency relation between the 
various spectral quantities, which can in principle be checked 
observationally, allowing one to distinguish between the cases of one, 
two, or more fields, provided that the observations are sufficiently accurate. 
We derived that consistency relation in subsection~\ref{obstestsec}. 
If the derivative of the spectral index can also be measured, there is an 
additional consistency relation, which was derived in the appendix.
With all these relations one should keep in mind that, while a large effect is
probably a clear proof of multiple-field effects being important, a small effect
might also signify the presence of trans-Planckian or short-distance physics
(see e.g.\ \cite{HuiKinney,SongKnox}).

An important ingredient of the calculation of the isocurvature amplitude was 
the observation that, although there may be many individual isocurvature 
perturbations in a multi-component system, only the total entropy perturbation 
$\tS$ enters into the equation of motion for the gravitational potential. 
With the assumptions that the various components behave as ideal fluids and 
have no interactions, the time derivative of $\tS$ on super-horizon scales during
radiation and matter domination was worked out in \eqref{tSder}. It was found 
that, if there are only two different types of energy in the universe (i.e.\ two
different equations of state, e.g.\ baryons and cold dark matter with $p=0$ and 
photons and hot dark matter with $p=\frac{1}{3}\gr$), $\tS$ is simply constant.
Although this is no longer true if there is a third type of energy, for example 
a cosmological constant with $p=-\gr$, we showed that for the general
three-component case the equation of motion for $\tS(t)$ only depends on the
background energy densities and can be computed explicitly, independently of the
gravitational potential. An explicit calculation of a three-component case with
matter, radiation, and a pure cosmological constant (with realistic values of 
the parameters) showed that the time evolution of $\tS$ before recombination 
can be completely neglected in that case.

To be more general we also derived, at the end of
subsection~\ref{scalarpertsec}, an  expression for the isocurvature
perturbation in the case that $\tS$ is not  constant. It turned out that the
effects of this can be absorbed in the  definition of the multiple-field term,
so that the general form of the  amplitudes and spectral indices is unchanged.
This means that the observational  relations in subsection~\ref{obstestsec} do
not depend on whether $\tS$ is constant or  not.

\ack

I thank Stefan Groot Nibbelink for useful discussions and comments. A
part of this work was done while I was still at the Spinoza Institute, Utrecht
University, The Netherlands.

\appendix
\section{Expressions for $\boldsymbol{\d \tn / \d \, \ln k}$}

The recent WMAP results have made it interesting to look at the derivatives of
the spectral indices with respect to $k$ as well. In a straightforward 
calculation (using \eqref{dereps} for the derivatives of the 
slow-roll functions), analogous to the derivation of the spectral indices
themselves in subsection~\ref{specquantinflsec}, we find to leading (i.e.\ 
second) order in slow roll:
\equa{
\fl
\left [ \dd[\tn_{\mathrm{ad}}]{\ln k} \right ]_{\mathrm{l.o.}}
= 2 \, \frac{4 \tge_\cH^2 + 3 \tge_\cH \tget^\parallel_\cH 
+ 3 (\tget^\parallel_\cH)^2 + 3 (\tget^\perp_\cH)^2 - \tgx^\parallel_\cH}
{1 + U_{P\, e}^T U_{P\, e}^{\,}}
\non \\
+ 2 \, \frac{2 e_2^T \lh - \tgx_{2\, \cH} + \tget^\perp_\cH \lh \tge_\cH 
+ 3 \tget^\parallel_\cH + 3 \gd_\cH \rh \rh U_{P\, e}^{\;}
+ U_{P\, e}^T \lh 2 \gd_\cH^2 - \frac{\dot{\gd}_\cH}{H_\cH} \rh U_{P\, e}^{\;}}
{1 + U_{P\, e}^T U_{P\, e}^{\,}}
\non \\
- 4 \lh \frac{2 \tge_\cH + \tget^\parallel_\cH 
+ 2 \tget^\perp_\cH e_2^T U_{P\, e}^{\;} + U_{P\, e}^T \gd_\cH U_{P\, e}^{\;}}
{1 + U_{P\, e}^T U_{P\, e}^{\,}} \rh^2,
\labl{dnad}\\
\fl
\left [ \dd[\tn_{\mathrm{iso}}]{\ln k} \right ]_{\mathrm{l.o.}}
= 2 \, \frac{V_e^T \lh 2 \gd_\cH^2 - \frac{\dot{\gd}_\cH}{H_\cH} \rh V_e}
{V_e^T V_e^{\,}}
- 4 \lh \frac{V_e^T \gd_\cH V_e^{\,}}{V_e^T V_e^{\,}} \rh^2,
\labl{dniso}\\
\fl
\left [ \dd[\tn_{\mathrm{mix}}]{\ln k} \right ]_{\mathrm{l.o.}}
= 2 \, \frac{e_2^T \lh - \tgx_{2\, \cH} + \tget^\perp_\cH \lh \tge_\cH 
+ 3 \tget^\parallel_\cH + 3 \gd_\cH \rh \rh V_e^{\;}
+ U_{P\, e}^T \lh 2 \gd_\cH^2 - \frac{\dot{\gd}_\cH}{H_\cH} \rh V_e^{\;}}
{U_{P\, e}^T V_e^{\;}}
\non\\
- 4 \lh \frac{\tget^\perp_\cH e_2^T V_e^{\,} + U_{P\, e}^T \gd_\cH V_e^{\,}}
{U_{P\, e}^T V_e^{\,}} \rh^2,
\labl{dnmix}\\
\fl
\left [ \dd[\tn_{\mathrm{tens}}]{\ln k} \right ]_{\mathrm{l.o.}}
= - 4 \tge_\cH \lh \tge_\cH + \tget^\parallel_\cH \rh.
\labl{dntens}
}
In these equations the derivative $\dot{\gd}_\cH \equiv \d\gd_\cH / \d t_\cH$ 
occurs. In specific models it can be calculated from the potential and the field
metric using the definitions \eqref{defdelta}:
\equ{
\fl
\frac{\dot{\gd}_\cH}{H_\cH} = 2 \tge_\cH \gd_\cH 
+ 2 \tge_\cH^{\;} \tget^\parallel_\cH \lh \Id + 2 e_1^{\;} e_1^T \rh
- \frac{\dot{\tM}{}^2_\cH}{3 H_\cH^3},
}
\equ{
\fl
\dot{\tM}{}^2_{mn} = (\cD_t \vc{e}_m) \cdot \tilde{\mx{M}}^2 \vc{e}_n
+ \vc{e}_m \cdot (\cD_t \tilde{\mx{M}}^2) \vc{e}_n
+ \vc{e}_m \cdot \tilde{\mx{M}}^2 (\cD_t \vc{e}_n),
}
\equ{
\fl
(\cD_t \tilde{\mx{M}}^2)^a_{\; b}
= G^{ac} (\nabla_b \nabla_c \nabla_e V) \dot{\gf}^e
- (\nabla_e R^a_{\; cdb}) \dot{\gf}^c \dot{\gf}^d \dot{\gf}^e
- 2 R^a_{\; cdb} (\cD_t \dot{\gf}^c) \dot{\gf}^d.
}
The derivatives of the basis vectors are given in equation \eqref{basisder}.

If we also include the $\d \tn_X / \d \, \ln k$ as observational quantities, 
there are, in the single-field case, two of those and one additional 
inflationary parameter ($\tgx^\parallel_\cH$), so that there should be one 
more consistency relation (in addition to \eqref{echteconsistrel1f}). It can 
be written as
\equ{
\frac{1}{\tn_{\mathrm{tens}}} \dd[\tn_{\mathrm{tens}}]{\ln k}
= \tn_{\mathrm{tens}} - \tn_{\mathrm{ad}}
\qquad\qquad\qquad\quad\;\;\:\mbox{(one field).}
}
In the two-field case, there are two more observational quantities but also
two more inflationary parameters ($\tgx_{2\, \cH}$ and $\dot{\gd}_\cH$), so that
we also have a single additional consistency relation:
\equ{
\fl
\frac{1}{\tn_{\mathrm{tens}}} \dd[\tn_{\mathrm{tens}}]{\ln k}
= \frac{\tn_{\mathrm{tens}} - \tn_{\mathrm{ad}} + r_{\mathrm{mix}}^2
\lh 2 \tn_{\mathrm{mix}} - \tn_{\mathrm{tens}} - \tn_{\mathrm{iso}} \rh}
{1 - r_{\mathrm{mix}}^2}
\qquad\mbox{(two fields),}
}
where $r_{\mathrm{mix}}^2$ is defined in \eqref{defrmix}.

\newcommand{\PRep}{{\it Phys.\ Rep.\ }}
\newcommand{\PTP}{{\it Prog.\ Theor.\ Phys.\ }}
\newcommand{\PTPS}{{\it Prog.\ Theor.\ Phys.\ Suppl.\ }}
\newcommand{\IJMP}{{\it Int.\ J.\ Mod.\ Phys.\ }}
\newcommand{\ApJS}{{\it Astrophys.\ J.\ Suppl.\ }}
\newcommand{\MNRAS}{{\it Mon.\ Not.\ R.\ Astron.\ Soc.\ }}
\newcommand{\JETPL}{{\it JETP\ Lett.\ }}


\section*{References}
\begin{thebibliography}{99}

\bibitem{Guth} Guth A H 1981 \PR D {\bf 23} 347
\bibitem{Starobinskymodel} Starobinsky A A 1980 \PL B {\bf 91} 99
\bibitem{boekLinde} Linde A D 1990 {\it Particle Physics and Inflationary
Cosmology} (Chur: Harwood Academic)
\bibitem{LythRiotto} Lyth D H and Riotto A 1999 \PRep {\bf 314} 1
({\it Preprint} hep-ph/9807278)
\bibitem{LiddleLythboek} Liddle A R and Lyth D H 2000 {\it Cosmological 
Inflation and Large-Scale Structure} (Cambridge: Cambridge University)
\bibitem{MAP} WMAP website: {\tt http://map.gsfc.nasa.gov}
\bibitem{WMAP} Bennett C L {\it et al} 2003 \ApJS {\bf 148} 1
({\it Preprint} astro-ph/0302207)
\bibitem{WMAPparam} Spergel D N {\it et al} 2003 \ApJS {\bf 148} 175
({\it Preprint} astro-ph/0302209)
\bibitem{WMAPinfl} Peiris H V {\it et al} 2003 \ApJS {\bf 148} 213
({\it Preprint} astro-ph/0302225)
\bibitem{GNvT2} Groot Nibbelink S and Van Tent B J W 2002 \CQG {\bf 19} 613
({\it Preprint} hep-ph/0107272)
\bibitem{Acquavivaetal} Acquaviva V, Bartolo N, Matarrese S and Riotto A 2003
\NP B {\bf 667} 119
({\it Preprint} astro-ph/0209156)
\bibitem{Habibetal} Habib S, Heitmann K, Jungman G and Molina-Paris C 2002
\PRL {\bf 89} 281301
({\it Preprint} astro-ph/0208443)
\bibitem{MartinSchwarzWKB} Martin J and Schwarz D J 2003 \PR D {\bf 67} 
083512
({\it Preprint} astro-ph/0210090)
\bibitem{LesgourguesPolarski} Lesgourgues J and Polarski D 1997 \PR D {\bf 56}
6425
({\it Preprint} astro-ph/9710083)
\bibitem{Bartoloetal} Bartolo N, Matarrese S and Riotto A 2001 \PR D {\bf 64}
123504
({\it Preprint} astro-ph/0107502)
\bibitem{Wandsetal2} Wands D, Bartolo N, Matarrese S and Riotto A 2002 \PR D
{\bf 66} 043520
({\it Preprint} astro-ph/0205253)
\bibitem{Tsujikawaetal} Tsujikawa S, Parkinson D and Bassett B A 2003 \PR D
{\bf 67} 083516
({\it Preprint} astro-ph/0210322)
\bibitem{BoyanovskyDeVega} Boyanovsky D and De Vega H J 2001 {\it Proc.\ 
7th Erice Chalonge School on Astrofundamental Physics}, ed N G Sanchez
(Dordrecht: Kluwer) p 37
({\it Preprint} astro-ph/0006446)
\bibitem{PolStargrav} Polarski D and Starobinsky A A 1995 \PL B {\bf 356} 196
({\it Preprint} astro-ph/9505125)
\bibitem{SasakiStewart} Sasaki M and Stewart E D 1996 \PTP {\bf 95} 71
({\it Preprint} astro-ph/9507001)
\bibitem{GarciaBWands} Garcia-Bellido J and Wands D 1996 \PR D {\bf 53} 5437
({\it Preprint} astro-ph/9511029) 
\bibitem{thesis} Van Tent B J W 2002 Cosmological Inflation with Multiple
Fields and the Theory of Density Fluctuations {\it PhD Thesis} Utrecht
University. Available on-line at\\ 
{\tt http://www.library.uu.nl/digiarchief/dip/diss/2002-1004-084000/inhoud.htm}
\bibitem{Mukhanovetal} Mukhanov V F, Feldman H A and Brandenberger R H 1992
\PRep {\bf 215} 203
\bibitem{Stewart} Stewart J M 1990 \CQG {\bf 7} 1169
\bibitem{Bardeen} Bardeen J M 1980 \PR D {\bf 22} 1882
\bibitem{Bassettetal} Bassett B A, Tamburini F, Kaiser D I and Maartens R 1999
\NP B {\bf 561} 188
({\it Preprint} hep-ph/9901319)
\bibitem{Wandsetal} Wands D, Malik K A, Lyth D H and Liddle A R 2000 \PR D
{\bf 62} 043527
({\it Preprint} astro-ph/0003278)
\bibitem{FinelliBrandenberger} Finelli F and Brandenberger R H 2000 \PR D 
{\bf 62} 083502
({\it Preprint} hep-ph/0003172)
\bibitem{thesisMalik} Malik K A 2001 Cosmological perturbations in an
inflationary universe {\it PhD Thesis} University of Portsmouth
({\it Preprint} astro-ph/0101563)
\bibitem{HenriquesMoorhouse} Henriques A B and Moorhouse R G 2002 \PR D 
{\bf 65} 103524
({\it Preprint} hep-ph/0109218)
\bibitem{TsujikawaBassett} Tsujikawa S and Bassett B A 2002 \PL B {\bf 536} 9
({\it Preprint} astro-ph/0204031)
\bibitem{TanakaBassett} Tanaka T and Bassett B A 2003 Application of the 
Separate Universe Approach to Preheating {\it Preprint} astro-ph/0302544
\bibitem{Starobinsky} Starobinsky A A 1979 \JETPL {\bf 30} 682
\bibitem{MartinSchwarz} Martin J and Schwarz D J 2000 \PR D {\bf 62} 103520
({\it Preprint} astro-ph/9911225)
\bibitem{PolarskiStarstates} Polarski D and Starobinsky A A 1996 \CQG {\bf 13}
377
({\it Preprint} gr-qc/9504030)
\bibitem{Kieferetal} Kiefer C, Polarski D and Starobinsky A A 1998 \IJMP D 
{\bf 7} 455
({\it Preprint} gr-qc/9802003)
\bibitem{LiddleLyth} Liddle A R and Lyth D H 1993 \PRep {\bf 231} 1
({\it Preprint} astro-ph/9303019)
\bibitem{BunnLiddleWhite} Bunn E F, Liddle A R and White M 1996 \PR D
{\bf 54} R5917
({\it Preprint} astro-ph/9607038)
\bibitem{vTGNcosm} Van Tent B J W and Groot Nibbelink S 2001 Inflationary
perturbations with multiple scalar fields {\it Preprint} hep-ph/0111370
\bibitem{Bargeretal} Barger V, Lee H-S and Marfatia D 2003 \PL B {\bf 565} 33
({\it Preprint} hep-ph/0302150)
\bibitem{Bridleetal} Bridle S L, Lewis A M, Weller J and Efstathiou G 2003
\MNRAS {\bf 342} L72
({\it Preprint} astro-ph/0302306)
\bibitem{Kinneyetal} Kinney W H, Kolb E W, Melchiorri A and Riotto A 2003 
WMAPping inflationary physics {\it Preprint} hep-ph/0305130
\bibitem{LeachLiddle} Leach S M and Liddle A R 2003 Constraining slow-roll
inflation with WMAP and 2dF {\it Preprint} astro-ph/0306305
\bibitem{Langlois} Langlois D 1999 \PR D {\bf 59} 123512
({\it Preprint} astro-ph/9906080)
\bibitem{Gordonetal} Gordon C, Wands D, Bassett B A and Maartens R 2001 \PR D
{\bf 63} 023506
({\it Preprint} astro-ph/0009131)
\bibitem{Bartoloetal1} Bartolo N, Matarrese S and Riotto A 2001 \PR D 
{\bf 64} 083514
({\it Preprint} astro-ph/0106022)
\bibitem{HwangNohSW} Hwang J-C and Noh H 1999 \PR D {\bf 59} 067302
({\it Preprint} astro-ph/9812007)
\bibitem{Hwangetal} Hwang J-C, Padmanabhan T, Lahav O and Noh H 2002 \PR D
{\bf 65} 043005
({\it Preprint} astro-ph/0107307)
\bibitem{Bucheretal} Bucher M, Moodley K and Turok N 2000 \PR D {\bf 62} 
083508
({\it Preprint} astro-ph/9904231)
\bibitem{Bucheretal2} Bucher M, Moodley K and Turok N 2002 \PR D {\bf 66}
023528
({\it Preprint} astro-ph/0007360)
\bibitem{Planck} Planck website: {\tt http://astro.estec.esa.nl/Planck}
\bibitem{HuiKinney} Hui L and Kinney W H 2002 \PR D {\bf 65} 103507
({\it Preprint} astro-ph/0109107)
\bibitem{Tanaka} Tanaka T 2000 A comment on trans-Planckian physics in
inflationary universe {\it Preprint} astro-ph/0012431
\bibitem{MartinBrandenberger} Martin J and Brandenberger R H 2003 \PR D 
{\bf 68} 063513
({\it Preprint} hep-th/0305161)
\bibitem{LythLiddle} Lyth D H and Liddle A R 1992 \PL B {\bf 291} 391
({\it Preprint} astro-ph/9208007)
\bibitem{Lidseyetal} Lidsey J E {\it et al} 1997 \RMP {\bf 69} 373
({\it Preprint} astro-ph/9508078)
\bibitem{SongKnox} Song Y-S and Knox L 2003 \PR D {\bf 68} 043518
({\it Preprint} astro-ph/0305411)
\bibitem{PolarskiStar} Polarski D and Starobinsky A A 1994 \PR D {\bf 50} 
6123
({\it Preprint} astro-ph/9404061)
\bibitem{KodamaSasaki} Kodama H and Sasaki M 1984 \PTPS {\bf 78} 1
\bibitem{HwangNoh} Hwang J-C and Noh H 2002 \CQG {\bf 19} 527
({\it Preprint} astro-ph/0103244)
\bibitem{Maliketal} Malik K A, Wands D and Ungarelli C 2003 \PR D {\bf 67} 
063516
({\it Preprint} astro-ph/0211602)

\end{thebibliography}
\end{document}